\documentstyle[aps,prl]{revtex}
\def\proof{\noindent{\bfseries Proof. }}

\begin{document}
\author{Lu-Ming Duan$^{1,2}$\thanks{%
Email: luming.duan@uibk.ac.at}, G. Giedke$^1$, J. I. Cirac$^1$, and P. Zoller%
$^1$}
\address{$^{1}$Institut f\"{u}r Theoretische Physik, Universit\"{a}t Innsbruck,
A-6020 Innsbruck, Austria \\
$^{2}$Laboratory of Quantum Communication and Quantum Computation,
University of Science and \\
Technology of China, Hefei 230026, China}
\title{Inseparability criterion for continuous variable systems}
\maketitle

\begin{abstract}
An inseparability criterion based on the total variance of a pair of
Einstein-Podolsky-Rosen type operators is proposed for continuous variable
systems. The criterion provides a sufficient condition for entanglement of
any two-party continuous variable states. Furthermore, for all the Gaussian
states, this criterion turns out to be a necessary and sufficient condition
for inseparability. %\newline

{\bf PACS numbers: }03.67.-a, 42.50.Dv, 89.70.+c
\end{abstract}

%\widetext

It is now believed that quantum entanglement plays an essential role in all
branches of quantum information theory \cite{1}. A problem of great
importance is then to check if a state, generally mixed, is entangled or
not. Concerning this problem, Peres proposed an inseparability criterion
based on partial transpose of the composite density operator \cite{2}, which
provides a sufficient condition for entanglement. This criterion was later
shown by Horodecki to be a necessary and sufficient condition for
inseparability of the $2\times 2$ or $2\times 3$ dimensional states, but not
to be necessary any more for higher dimensional states \cite{3,4}. Many
recent protocols for quantum communication and computation are based on
continuous variable quantum systems \cite{5,6,7,8,9,10,11}, and the
continuous variable optical system has been used to experimentally realize
the unconditional quantum teleportation \cite{12}. Hence, it is desirable to
know if a continuous variable state is entangled or not.

In this paper, we propose a simple inseparability criterion for continuous
variable states. The criterion is based on the calculation of the total
variance of a pair of Einstein-Podolsky-Rosen (EPR) type operators. We find
that for any separable continuous variable states, the total variance is
bounded from below by a certain value resulting from the uncertainty
relation, whereas for entangled states this bound can be exceeded. So
violation of this bound provides a sufficient condition for inseparability
of the state. Then we investigate how strong the bound is for the set of
Gaussian states, which are of great practical importance. It is shown that
for a Gaussian state, the compliance with the low bound by a certain pair of
EPR type operators guarantees that the state has a P-representation with
positive distribution, so the state must be separable. Hence we obtain a
necessary and sufficient inseparability criterion for all the Gaussian
continuous variable states.

We say a quantum state $\rho $ of two modes $1$ and $2$ is separable if and
only if it can be expressed in the following form

\begin{equation}  \label{e1}
\rho =\mathrel{\mathop{\sum }\limits_{i}}p_{i}\rho _{i1}\otimes \rho _{i2},
\end{equation}
where we assume $\rho _{i1}$ and $\rho _{i2}$ to be normalized states of the
modes $1$ and $2$, respectively, and $p_{i}\geq 0$ to satisfy $%
\mathrel{\mathop{\sum }\limits_{i}}p_{i}=1$.

A maximally entangled continuous variable state can be expressed as a
co-eigenstate of a pair of EPR-type operators \cite{13}, such as $\widehat{x}%
_{1}+\widehat{x}_{2}$ and $\widehat{p}_{1}-\widehat{p}_{2}$. So the total
variance of these two operators reduces to zero for maximally entangled
continuous variable states. Of course, the maximally entangled continuous
variable states are not physical, but for the physical entangled continuous
variable states--the two-mode squeezed states \cite{14}, this variance will
rapidly tend to zero by increasing the degree of squeezing. Interestingly,
we find that for any separable state, there exists a lower bound to the
total variance. To be more general, we consider the following type of
EPR-like operators:

\begin{mathletters}
\label{e2}
\begin{eqnarray}
\widehat{u} &=&|a|\widehat{x}_{1}+\frac{1}{a}\widehat{x}_{2},  \label{2a} \\
\widehat{v} &=&|a|\widehat{p}_{1}-\frac{1}{a}\widehat{p}_{2},  \label{2b}
\end{eqnarray}
where we assume $a$ is an arbitrary (nonzero) real number. For any separable
state, the total variance of any pair of EPR-like operators in the form of
Eq. (2) should satisfy a lower bound indicated by the following theorem:

{\bf Theorem 1 (sufficient criterion for inseparability):} For any separable
quantum state $\rho $, the total variance of a pair of EPR-like operators
defined by Eq. (2) with the commutators $\left[ \widehat{x}_{j},\widehat{p}%
_{j^{^{\prime }}}\right] =i\delta _{jj^{^{\prime }}}$ $\left( j,j^{^{\prime
}}=1,2\right) $ satisfies the inequality

\end{mathletters}
\begin{equation}  \label{e3}
\left\langle \left( \Delta \widehat{u}\right) ^{2}\right\rangle _{\rho
}+\left\langle \left( \Delta \widehat{v}\right) ^{2}\right\rangle _{\rho
}\geq a^{2}+\frac{1}{a^{2}}.
\end{equation}

{\bf Proof.} We can directly calculate the total variance of the $\widehat{u}
$ and $\widehat{v}$ operators using the decomposition (1) of the density
operator $\rho $, and finally get the following expression:

\begin{eqnarray}
&&\left\langle \left( \Delta \widehat{u}\right) ^{2}\right\rangle _{\rho
}+\left\langle \left( \Delta \widehat{v}\right) ^{2}\right\rangle _{\rho }=%
%TCIMACRO{\underset{i}{\sum }}%
%BeginExpansion
\mathrel{\mathop{\sum }\limits_{i}}%
%EndExpansion
p_{i}\left( \left\langle \widehat{u}^{2}\right\rangle _{i}+\left\langle 
\widehat{v}^{2}\right\rangle _{i}\right) -\left\langle \widehat{u}%
\right\rangle _{\rho }^{2}-\left\langle \widehat{v}\right\rangle _{\rho }^{2}
\nonumber \\
&=&%
%TCIMACRO{\underset{i}{\sum }}%
%BeginExpansion
\mathrel{\mathop{\sum }\limits_{i}}%
%EndExpansion
p_{i}\left( a^{2}\left\langle \widehat{x}_{1}^{2}\right\rangle _{i}+\frac{1}{%
a^{2}}\left\langle \widehat{x}_{2}^{2}\right\rangle _{i}+a^{2}\left\langle 
\widehat{p}_{1}^{2}\right\rangle _{i}+\frac{1}{a^{2}}\left\langle \widehat{p}%
_{2}^{2}\right\rangle _{i}\right)  \nonumber \\
&&+2\frac{a}{\left| a\right| }\left( 
%TCIMACRO{\underset{i}{\sum }}%
%BeginExpansion
\mathrel{\mathop{\sum }\limits_{i}}%
%EndExpansion
p_{i}\left\langle \widehat{x}_{1}\right\rangle _{i}\left\langle \widehat{x}%
_{2}\right\rangle _{i}-%
%TCIMACRO{\underset{i}{\sum }}%
%BeginExpansion
\mathrel{\mathop{\sum }\limits_{i}}%
%EndExpansion
p_{i}\left\langle \widehat{p}_{1}\right\rangle _{i}\left\langle \widehat{p}%
_{2}\right\rangle _{i}\right) -\left\langle \widehat{u}\right\rangle _{\rho
}^{2}-\left\langle \widehat{v}\right\rangle _{\rho }^{2}  \label{e4} \\
&=&\mathrel{\mathop{\sum }\limits_{i}}p_{i}\left( a^{2}\left\langle \left(
\Delta \widehat{x}_{1}\right) ^{2}\right\rangle _{i}+\frac{1}{a^{2}}%
\left\langle \left( \Delta \widehat{x}_{2}\right) ^{2}\right\rangle
_{i}+a^{2}\left\langle \left( \Delta \widehat{p}_{1}\right)
^{2}\right\rangle _{i}+\frac{1}{a^{2}}\left\langle \left( \Delta \widehat{p}%
_{2}\right) ^{2}\right\rangle _{i}\right)  \nonumber \\
&&+\mathrel{\mathop{\sum }\limits_{i}}p_{i}\left\langle \widehat{u}%
\right\rangle _{i}^{2}-\left( \mathrel{\mathop{\sum }\limits_{i}}%
p_{i}\left\langle \widehat{u}\right\rangle _{i}\right) ^{2}+%
\mathrel{\mathop{\sum }\limits_{i}}p_{i}\left\langle \widehat{v}%
\right\rangle _{i}^{2}-\left( \mathrel{\mathop{\sum }\limits_{i}}%
p_{i}\left\langle \widehat{v}\right\rangle _{i}\right) ^{2}.  \nonumber
\end{eqnarray}
In Eq. (4), the symbol $\left\langle \cdots \right\rangle _{i}$ denotes
average over the product density operator $\rho _{i1}\otimes \rho _{i2}$. It
follows from the uncertainty relation that $\left\langle \left( \Delta 
\widehat{x}_{j}\right) ^{2}\right\rangle _{i}+\left\langle \left( \Delta 
\widehat{p}_{j}\right) ^{2}\right\rangle _{i}\geq \left| \left[ \widehat{x}%
_{j},\widehat{p}_{j}\right] \right| =1$ for $j=1,2$, and moreover, by
applying the Cauchy-Schwarz inequality $\left( 
\mathrel{\mathop{\sum
}\limits_{i}}p_{i}\right) \left( \mathrel{\mathop{\sum }\limits_{i}}%
p_{i}\left\langle \widehat{u}\right\rangle _{i}^{2}\right) \geq \left( 
\mathrel{\mathop{\sum
}\limits_{i}}p_{i}\left| \left\langle \widehat{u}\right\rangle _{i}\right|
\right) ^{2},$ we know that the last line of Eq. (4) is bounded from below
by zero. Hence, the total variance of the two EPR-like operators $\widehat{u}
$ and $\widehat{v}$ is bounded from below by $a^{2}+\frac{1}{a^{2}}$ for any
separable state. This completes the proof of the theorem.$\Box$

Note that this theorem in fact gives a set of inequalities for separable
states. The operators $\widehat{x}_{j},\widehat{p}_{j}$ $\left( j=1,2\right) 
$ in the definition (1) can be any local operators satisfying the
commutators $\left[ \widehat{x}_{j},\widehat{p}_{j^{^{\prime }}}\right]
=i\delta _{jj^{^{\prime }}}$. In particular, if we apply an arbitrary local
unitary operation $U_{1}\otimes U_{2}$ to the operators $\widehat{u}$ and $%
\widehat{v}$, the inequality (3) remains unchanged. Note also that without
loss of generality we have taken the operators $x_{j}$ and $p_{j}$
dimensionless.

For inseparable states, the total variance of the $\widehat{u}$ and $%
\widehat{v}$ operators is required by the uncertainty relation to be larger
than or equal to $\left| a^{2}-\frac{1}{a^{2}}\right| $ , which reduces to
zero for $a=1$. For separable states the much stronger bound given by Eq.
(3) must be satisfied. A natural question is then how strong the bound is.
Is it strong enough to ensure that if some inequality in the form of Eq. (3)
is satisfied, the state necessarily becomes separable? Of course, it will be
very difficult to consider this problem for arbitrary continuous variable
states. However, in recent experiments and protocols for quantum
communication \cite{5,6,7,8,9,10,11,12}, continuous variable entanglement is
generated by two-mode squeezing or by beam splitters, and the communication
noise results from photon absorption and thermal photon emission. All these
processes lead to Gaussian states. So, we will limit ourselves to consider
Gaussian states, which are of great practical importance. We find that the
inequality (3) indeed gives a necessary and sufficient inseparability
criterion for all the Gaussian states. To present and prove our main
theorem, we need first mention some notations and results for Gaussian
states.

It is convenient to represent a Gaussian state by its Wigner characteristic
function. A two-mode state with the density operator $\rho $ has the
following Wigner characteristic function \cite{14}

\begin{eqnarray}
\chi ^{\left( w\right) }\left( \lambda _{1},\lambda _{2}\right) &=&tr\left[
\rho \exp \left( \lambda _{1}\widehat{a}_{1}-\lambda _{1}^{\ast }\widehat{a}%
_{1}^{\dagger }+\lambda _{2}\widehat{a}_{2}-\lambda _{2}^{\ast }\widehat{a}%
_{2}^{\dagger }\right) \right]  \nonumber \\
&=&tr\left\{ \rho \exp \left[ i\sqrt{2}\left( \lambda _{1}^{I}\widehat{x}%
_{1}+\lambda _{1}^{R}\widehat{p}_{1}+\lambda _{2}^{I}\widehat{x}_{2}+\lambda
_{2}^{R}\widehat{p}_{2}\right) \right] \right\} ,  \label{e5}
\end{eqnarray}
where the parameters $\lambda _{j}=\lambda _{j}^{R}+i\lambda _{j}^{I}$, and
the annihilation operators $\widehat{a}_{j}=\frac{1}{\sqrt{2}}\left( 
\widehat{x}_{j}+i\widehat{p}_{j}\right) $, with the quadrature amplitudes $%
\widehat{x}_{j},\widehat{p}_{j}$ satisfying the commutators $\left[ \widehat{%
x}_{j},\widehat{p}_{j^{^{\prime }}}\right] =i\delta _{jj^{^{\prime }}}$ $%
\left( j,j^{^{\prime }}=1,2\right) $. For a Gaussian state, the Wigner
characteristic function $\chi ^{\left( w\right) }\left( \lambda _{1},\lambda
_{2}\right) $ is a Gaussian function of $\lambda _{j}^{R}$ and $\lambda
_{j}^{I}$ \cite{14}. Without loss of generality, we can write $\chi ^{\left(
w\right) }\left( \lambda _{1},\lambda _{2}\right) $ in the form

\begin{equation}  \label{e6}
\chi ^{\left( w\right) }\left( \lambda _{1},\lambda _{2}\right) =\exp \left[
-\frac{1}{2}\left( \lambda _{1}^{I},\lambda _{1}^{R},\lambda
_{2}^{I},\lambda _{2}^{R}\right) M\left( \lambda _{1}^{I},\lambda
_{1}^{R},\lambda _{2}^{I},\lambda _{2}^{R}\right) ^{T}\right]  \label{e6}
\end{equation}
In Eq. (6), linear terms in the exponent are not included since they can be
easily removed by some local displacements of $\widehat{x}_{j},\widehat{p}%
_{j}$ and thus have no influence on separability or inseparability of the
state. The correlation property of the Gaussian state is completely
determined by the $4\times 4$ real symmetric correlation matrix $M$, which
can be expressed as

\begin{equation}  \label{e7}
M=\left( 
\begin{array}{ll}
G_{1} & C \\ 
C^{T} & G_{2}
\end{array}
\right) ,
\end{equation}
where $G_{1},$ $G_{2},$ and $C$ are $2\times 2$ real matrices. To study the
separability property, it is convenient to first transform the Gaussian
state to some standard forms through local linear unitary Bogoliubov
operations (LLUBOs) $U_{l}=U_{1}\otimes U_{2}$. In the Heisenberg picture,
the general form of the LLUBO $U_{l}$ is expressed as $U_{l}\left( \widehat{x%
}_{j},\widehat{p}_{j}\right) ^{T}U_{l}^{\dagger }=H_{j}\left( \widehat{x}%
_{j},\widehat{p}_{j}\right) ^{T}$ for $j=1,2$ , where $H_{j}$ is some $%
2\times 2$ real matrix with $\det H_{j}=1.$ Any LLUBO is obtainable by
combining the squeezing transformation together with some rotations \cite{15}%
. We have the following two lemmas concerning the standard forms of the
Gaussian state:

{\bf Lemma 1 (standard form I)}: Any Gaussian state $\rho _G$ can be
transformed through LLUBOs to the standard form I with the correlation
matrix given by

\begin{equation}  \label{e8}
M_s^I=\left( 
\begin{array}{llll}
n &  & c &  \\ 
& n &  & c^{^{\prime }} \\ 
c &  & m &  \\ 
& c^{\prime } &  & m
\end{array}
\right) ,\text{ }\left( n,m\geq 1\right)
\end{equation}

\noindent {\bf Proof.}{\ } A LLUBO on the state $\rho _{G}$ transforms the
correlation matrix $M$ in the Wigner characteristic function in the
following way 
\begin{equation}  \label{e9}
\left( 
\begin{array}{ll}
V_{1} &  \\ 
& V_{2}
\end{array}
\right) M\left( 
\begin{array}{ll}
V_{1}^{T} &  \\ 
& V_{2}^{T}
\end{array}
\right) ,
\end{equation}
where $V_{1}$ and $V_{2}$ are real matrices with $\det V_{1}=\det V_{2}=1$.
Since the matrices $G_{1}$ and $G_{2}$ in Eq. (7) are real symmetric, we can
choose first a LLUBO with orthogonal $V_{1}$ and $V_{2}$ which diagonalize $%
G_{1}$ and $G_{2}$, and then a local squeezing operation which transforms
the diagonalized $G_{1}$ and $G_{2}$ into the matrices $G_{1}^{^{\prime
}}=nI_{2}$ and $G_{2}^{^{\prime }}=mI_{2}$, respectively, where $I_{2}$ is
the $2\times 2$ unit matrix. After these two steps of operations, we assume
the matrix $C$ in Eq. (7) is changed into $C^{^{\prime }}$, which always has
a singular value decomposition, thus it can be diagonalized by another LLUBO
with suitable orthogonal $V_{1}$ and $V_{2}$. The last orthogonal LLUBO does
not influence $G_{1}^{^{\prime }}$ and $G_{2}^{^{\prime }}$ any more since
they are proportional to the unit matrix. Hence, any Gaussian state can be
transformed by three-step LLUBOs to the standard form I. The four parameters 
$n,m,c,$ and $c^{^{\prime }}$ in the standard form I are related to the four
invariants $\det G_{1},$ $\det G_{2}$, $\det C$, and $\det M$ of the
correlation matrix under LLUBOs by the equations $\det G_{1}=n^{2},$ $\det
G_{2}=m^{2},$ $\det C=cc^{^{\prime }},$ and $\det M=\left( nm-c^{2}\right)
\left( nm-c^{^{\prime }2}\right)$.$\Box$

{\bf Lemma 2 (standard form II):} Any Gaussian state $\rho _{G}$ can be
transformed through LLUBOs into the standard form II with the correlation
matrix given by

\begin{equation}  \label{e10}
M_{s}^{II}=\left( 
\begin{array}{llll}
n_{1} &  & c_{1} &  \\ 
& n_{2} &  & c_{2} \\ 
c_{1} &  & m_{1} &  \\ 
& c_{2} &  & m_{2}
\end{array}
\right) ,
\end{equation}
where the $n_{i},$ $m_{i}$ and $c_{i}$ satisfy

\begin{mathletters}
\label{e11}
\begin{eqnarray}
\frac{n_{1}-1}{m_{1}-1}& =&\frac{n_{2}-1}{m_{2}-1},  \label{11a} \\
\left| c_{1}\right| -\left| c_{2}\right| &=& \sqrt{\left( n_{1}-1\right)
\left( m_{1}-1\right) }-\sqrt{\left( n_{2}-1\right) \left( m_{2}-1\right) }.
\label{11b}
\end{eqnarray}
\noindent {\bf Proof.}{\ } First, any Gaussian state can be tranformed
through LLUBOs to the standard form I. We then apply two additional local
squeezing operations on the standard form I, and get the state with the
following correlation matrix

\end{mathletters}
\begin{equation}  \label{e12}
M^{^{\prime }}=\left( 
\begin{array}{llll}
nr_{1} &  & \sqrt{r_{1}r_{2}}c &  \\ 
& \frac{n}{r_{1}} &  & \frac{c^{^{\prime }}}{\sqrt{r_{1}r_{2}}} \\ 
\sqrt{r_{1}r_{2}}c &  & mr_{2} &  \\ 
& \frac{c^{^{\prime }}}{\sqrt{r_{1}r_{2}}} &  & \frac{m}{r_{2}}
\end{array}
\right) ,
\end{equation}
where $r_{1}$ and $r_{2}$ are arbitrary squeezing parameters. $M^{^{\prime
}} $ in Eq. (12) has the standard form $M_{s}^{II}$ (10) if $r_{1}$ and $%
r_{2}$ satisfy the following two equations 
\begin{equation}  \label{e13}
\frac{\frac{n}{r_{1}}-1}{nr_{1}-1}=\frac{\frac{m}{r_{2}}-1}{mr_{2}-1},
\end{equation}
\begin{equation}  \label{e14}
\sqrt{r_{1}r_{2}}\left| c\right| -\frac{\left| c^{^{\prime }}\right| }{\sqrt{%
r_{1}r_{2}}}=\sqrt{\left( nr_{1}-1\right) \left( mr_{2}-1\right) }-\sqrt{%
\left( \frac{n}{r_{1}}-1\right) \left( \frac{m}{r_{2}}-1\right) }.
\end{equation}
Our task remains to prove that Eqs. (13) and (14) are indeed satisfied by
some positive $r_{1}$ and $r_{2}$ for arbitrary Gaussian states$.$ Without
loss of generality, we assume $\left| c\right| \geq \left| c^{^{\prime
}}\right| $ and $n\geq m$. From Eq. (13), $r_{2}$ can be expressed as a
continuous function of $r_{1}$ with $r_{2}\left( r_{1}=1\right) =1$ and $%
r_{2}\left( r_{1}\right) \stackrel{r_{1}\rightarrow \infty }{\longrightarrow 
}m.$ Substituting this expression $r_{2}\left( r_{1}\right) $ into Eq. (14),
we construct a function $f\left( r_{1}\right) $ by subtracting the right
hand side of Eq. (14) from the left hand side, i.e., $f\left( r_{1}\right) =$%
Left$(14)-$Right$(14)$. Obviously, $f\left( r_{1}=1\right) =\left| c\right|
-\left| c^{^{\prime }}\right| \geq 0,$ and $f\left( r_{1}\right) \stackrel{%
r_{1}\rightarrow \infty }{\longrightarrow }\sqrt{r_{1}m}\left( \left|
c\right| -\sqrt{n\left( m-\frac{1}{m}\right) }\right) \leq 0,$ where the
inequality $\left| c\right| \leq \sqrt{n\left( m-\frac{1}{m}\right) }$
results from the physical condition $\left\langle \left( \Delta \widehat{u}%
_{0}\right) ^{2}\right\rangle +\left\langle \left( \Delta \widehat{v}%
_{0}\right) ^{2}\right\rangle \geq \left| \left[ \widehat{u}_{0},\widehat{v}%
_{0}\right] \right| $ with $\widehat{u}_{0}=\sqrt{m-\frac{1}{m}}\widehat{x}%
_{1}-\frac{c}{\left| c\right| }\sqrt{n}\widehat{x}_{2}$ and $\widehat{v}_{0}=%
\frac{\sqrt{n}}{m}\widehat{p}_{2}.$ It follows from continuity that there
must exist a $r_{1}^{\ast }\in \left[ 1,\infty \right) $ which makes $%
f\left( r_{1}=r_{1}^{\ast }\right) =0.$ So Eqs. (13) and (14) have at least
one solution. This proves lemma 2.$\Box$

We remark that corresponding to a given standard form I or II, there are a
class of Gaussian states, which are equivalent under LLUBOs. Note that
separability or inseparability is a property not influenced by LLUBOs, so
all the Gaussian states with the same standard forms have the same
separability or inseparability property. With the above preparations, now we
present the following main theorem:

{\bf Theorem 2 (necessary and sufficient inseparability criterion for
Gaussian states):} A Gaussian state $\rho _{G}$ is separable if and only if
when expressed in its standard form II, the inequality (3) is satisfied by
the following two EPR-type operators 
\begin{mathletters}
\label{e15}
\begin{eqnarray}
\widehat{u} &=&a_{0}\widehat{x}_{1}-\frac{c_{1}}{\left| c_{1}\right| }\frac{1%
}{a_{0}}\widehat{x}_{2},  \label{15a} \\
\widehat{v} &=&a_{0}\widehat{p}_{1}-\frac{c_{2}}{\left| c_{2}\right| }\frac{1%
}{a_{0}}\widehat{p}_{2},  \label{15b}
\end{eqnarray}
where $a_{0}^{2}=\sqrt{\frac{m_{1}-1}{n_{1}-1}}=\sqrt{\frac{m_{2}-1}{n_{2}-1}%
}.$

%TCIMACRO{
%\TeXButton{Proof}{\proof%
%}}%
%BeginExpansion
\proof%
%
%EndExpansion
The `only if' part follows directly from theorem 1. We only need to prove
the `if' part. From lemma 2, we can first transform the Gaussian state
through LLUBOs to the standard form II. The state after transformation is
denoted by $\rho _{G}^{II}$. Then, substituting the expression (15) of $%
\widehat{u}$ and $\widehat{v}$ into the inequality (3), and calculating $%
\left\langle \left( \Delta \widehat{u}\right) ^{2}\right\rangle
+\left\langle \left( \Delta \widehat{v}\right) ^{2}\right\rangle $ using the
correlation matrix $M_{s}^{II},$ we get the following inequality

\end{mathletters}
\begin{equation}  \label{e16}
a_{0}^{2}\frac{n_{1}+n_{2}}{2}+\frac{m_{1}+m_{2}}{2a_{0}^{2}}-\left|
c_{1}\right| -\left| c_{2}\right| \geq a_{0}^{2}+\frac{1}{a_{0}^{2}},
\end{equation}
which, combined with Eq. (11), yields

\begin{mathletters}
\label{e17}
\begin{eqnarray}
\left| c_{1}\right| &\leq &\sqrt{\left( n_{1}-1\right) \left( m_{1}-1\right) 
}.  \label{17a} \\
\left| c_{2}\right| &\leq &\sqrt{\left( n_{2}-1\right) \left( m_{2}-1\right) 
}  \label{17b}
\end{eqnarray}
The inequality (17) ensures that the matrix $M_{s}^{^{II}}-I$ is positive
semi-definite. So there exists a Fourier transformation to the following
normal characteristic function of the state $\rho _{G}^{II}$ 
\end{mathletters}
\begin{eqnarray}
\chi _{II}^{\left( n\right) }\left( \lambda _{1},\lambda _{2}\right) &=&\chi
_{II}^{\left( w\right) }\left( \lambda _{1},\lambda _{2}\right) \exp \left[ 
\frac{1}{2}\left( \left| \lambda _{1}\right| ^{2}+\left| \lambda _{2}\right|
^{2}\right) \right]  \nonumber \\
&=&\exp \left[ -\frac{1}{2}\left( \lambda _{1}^{I},\lambda _{1}^{R},\lambda
_{2}^{I},\lambda _{2}^{R}\right) \left( M_{s}^{^{II}}-I\right) \left(
\lambda _{1}^{I},\lambda _{1}^{R},\lambda _{2}^{I},\lambda _{2}^{R}\right)
^{T}\right].  \label{e18}
\end{eqnarray}
This means that $\rho _{G}^{II}$ can be expressed as 
\begin{equation}  \label{e19}
\rho _{G}^{II}=\int d^{2}\alpha d^{2}\beta P\left( \alpha ,\beta \right)
\left| \alpha ,\beta \right\rangle \left\langle \alpha ,\beta \right| ,
\end{equation}
where $P\left( \alpha ,\beta \right) $ is the Fourier transformation of $%
\chi _{II}^{\left( n\right) }\left( \lambda _{1},\lambda _{2}\right) $ and
thus is a positive Gaussian function. Eq. (19) shows $\rho _{G}^{II}$ is
separable. Since the original Gaussian state $\rho _{G}$ differs from $\rho
_{G}^{II}$ by only some LLUBOs, it must also be separable. This completes
the proof of theorem 2.$\Box$

Now we have a necessary and sufficient inseparability criterion for all the
Gaussian states. We conclude the paper by applying this criterion to a
simple example. Consider a two-mode squeezed vacuum state $e^{-r\left( 
\widehat{a}_{1}^{\dagger }\widehat{a}_{2}^{\dagger }-\widehat{a}_{1}\widehat{%
a}_{2}\right) }\left| \text{vac}\right\rangle $ with the squeezing parameter 
$r$. This state has been used in recent experiment for continuous variable
quantum teleportation \cite{12}. Suppose that the two optical modes are
subject to independent thermal noise during transmission with the same
damping coefficient denoted by $\eta $ and the same mean thermal photon
number denoted by $\overline{n}.$ It is easy to show that after time $t$,
the standard correlation matrix for this Gaussian state has the form of Eq.
(8) with $n=m=\cosh \left( 2r\right) e^{-2\eta t}+\left( 2\overline{n}%
+1\right) \left( 1-e^{-2\eta t}\right) $ and $c=-c^{^{\prime }}=\sinh \left(
2r\right) e^{-2\eta t}$ \cite{16}. So the inseparability criterion means
that if the transmission time $t$ satisfies 
\begin{equation}  \label{e20}
t<\frac{1}{2\eta }\ln \left( 1+\frac{1-e^{-2r}}{2\overline{n}}\right) ,
\end{equation}
the state is entangled; otherwise it becomes separable. Interestingly. Eq.
(20) shows that if there is only vacuum fluctuation noise, i.e., $\overline{n%
}=0$ (this seems to be a good approximation for optical frequency), the
initial squeezed state is always entangled. This result does not remain true
if thermal noise is present. In the limit $\overline{n}\gg 1$, the state is
not entangled any more when the transmission time $t\geq \frac{1-e^{-2r}}{%
4\eta \overline{n}}$.

{\bf Note added:} After submission of this work, we became aware of a recent
preprint by R. Simon (quant-ph/9909044), which shows that the
Peres-Horodecki criterion also provides a necessary and sufficient condition
for inseparability of Gaussian continuous variable quantum states.

{\bf Acknowledgments}

This work was funded by the Austrian Science Foundation and by the European
TMR network Quantum Information. GG acknowledges support by the
Friedrich-Naumann-Stiftung. %\newpage

\end{document}